\documentclass[12pt,preprint]{aastex}

\usepackage{amsmath}
\usepackage{amssymb}
\usepackage{latexsym}
\usepackage{graphicx}

\newcommand{\myemail}{fabio.mattana@apc.univ-paris7.fr.}

\newcommand{\PWN}{PWN}
\newcommand{\PWNe}{PWNe}

\newcommand{\hess}{\textit{H.E.S.S.}}
\newcommand{\integral}{\textit{INTEGRAL}}

\newcommand{\xmm}{\textit{XMM-Newton}}
\newcommand{\chandra}{\textit{Chandra}}
\newcommand{\beppo}{Beppo\textit{SAX}}
\newcommand{\rosat}{\textit{ROSAT}}
\newcommand{\suzaku}{\textit{Suzaku}}

\renewcommand{\deg}{${^\circ}$}

\newcommand{\flux}{\mbox{erg cm$^{-2}$ s$^{-1}$}}

\newcommand{\lum}{\mbox{erg s$^{-1}$}}

\newcommand{\ki}{$\chi^2$}
\newcommand{\kir}{$\chi^2_{red}$}

\slugcomment{{\sc Submitted to ApJ}}

\shorttitle{Extended hard X-ray emission from the Vela PWN}
\shortauthors{Mattana et al.}

\begin{document}


\title{Extended hard X-ray emission from the Vela pulsar wind nebula}


\author{F. Mattana\altaffilmark{1}, D. G\"otz\altaffilmark{2}, R. Terrier\altaffilmark{1}, L. Bouchet\altaffilmark{3}, G. Ponti\altaffilmark{4}, M. Falanga\altaffilmark{5}, M. Renaud\altaffilmark{6}, \\
I. Caballero\altaffilmark{2}, S. Soldi\altaffilmark{2}, J.~A. Zurita Heras\altaffilmark{1}, and S. Schanne\altaffilmark{2}}


\altaffiltext{1}{Fran\c{c}ois Arago Centre, APC (UMR 7164 Universit\'e Paris Diderot, CNRS/IN2P3, CEA/DSM, Observatoire de Paris), 13 rue Watt, F-75205 Paris cedex 13, France; \myemail}
\altaffiltext{2}{AIM (UMR 7158 CEA/DSM - CNRS - Universit\'e Paris Diderot) Irfu/Service d'Astrophysique, F-91191 Gif-sur-Yvette, France}
\altaffiltext{3}{Universit\'e de Toulouse, UPS-OMP, IRAP,  Toulouse, France
CNRS, IRAP, 9 Av. colonel Roche, BP 44346, F-31028 Toulouse cedex 4, France}
\altaffiltext{4}{Faculty of Physical and Applied Science, University of Southampton, Southampton, SO17 1BJ, UK}
\altaffiltext{5}{International Space Science Institute (ISSI), Hallerstrasse 6, CH-3012 Bern, Switzerland}
\altaffiltext{6}{Laboratoire Univers et Particules de Montpellier (LUPM), Universit\'e Montpellier II, CNRS/IN2P3 UMR 5299, F-34095 Montpellier, France}


\begin{abstract}
The nebula powered by the Vela pulsar is one of the best examples of an evolved pulsar wind nebula, allowing to access  the particle injection history and the interaction with the supernova ejecta. We report on the \integral\ discovery of extended emission above 18 keV from the Vela nebula. The northern side has no known counterparts and it appears larger and more significant than the southern one, which is in turn partially coincident with the cocoon, the soft X-ray and TeV filament towards the centre of the remnant. We also present the spectrum of the Vela nebula in the 18--400 keV energy range as measured by IBIS/ISGRI and SPI onboard the \integral\ satellite. The apparent discrepancy between IBIS/ISGRI, SPI, and previous measurements is understood in terms of point spread function, supporting the hypothesis of a nebula more diffuse than previously thought. A break at  $\sim$25 keV is found in the spectrum within 6\arcmin\ from the pulsar after including the \suzaku\ XIS data. Interpreted as a cooling break, this points out that the inner nebula is composed by electrons injected in the last $\sim$2000 years. Broad-band modeling also implies a magnetic field higher than 10 $\mu$G in this region. Finally, we discuss the nature of the northern emission, which might be due to fresh particles injected after the passage of the reverse shock.
\end{abstract}

\keywords{pulsars: general --- pulsars: individual (PSR B0833--45) --- ISM: supernova remnants --- ISM: individual objects (Vela PWN) --- X-rays: general}


\section{Introduction}
\label{sec:1}

Pulsar Wind Nebulae (\PWNe) are the non-thermal bubbles inflated by the winds of rotation-powered pulsars. Recent observations 
have allowed to arrange the variety of their morphologies in an evolutionary sequence resulting from the interaction with their surroundings \citep{GaenslerSlane06}. 
A very complex phase occurs after the host remnant evolves into the Sedov-Taylor phase ($\sim$10 kyr after the pulsar birth), when a reverse shock propagates inwards into the supernova ejecta and eventually collides with the \PWN\ \citep{vanderswaluw01}. Located at a distance of 290 pc \citep{Dodson03a}, the \PWN\ powered by the Vela pulsar \citep[PSR B0833--45, with spin-down luminosity  $\dot{E} = 6.9 \times10^{36}$ \lum, and characteristic age $\sim$11.4 kyr;][]{Dodson07}  is the best example of a \PWN\ in such a phase, which is thought to be typical of composite supernova remnants.
 
The prime evidence of the pulsar wind is the structured \PWN\ resolved by \chandra\ within $\sim$1\arcmin\ from the pulsar  \citep{Helfand01,Pavlov01,Pavlov03}. The X-ray flux measured by \citet{Mangano05} at varying integration radii shows that the \chandra\ \PWN\ is surrounded by fainter emission up to at least 15\arcmin. At a larger scale, {\em ROSAT} found a X-ray filament (the so-called cocoon) extending $\sim$45\arcmin\  south/south-west  from the pulsar \citep{Markwardt95}. The cocoon also stands out among the radio filaments composing the large  ($\sim$1.5\deg\ radius) Vela X nebula \citep{Rishbeth58}. All these structures are embedded in the 
Vela supernova remnant, which has a radius of $\sim$4$^\circ$. 

The cocoon is ascribed to an asymmetric reverse shock, which stripped the bulk of the particles from around the pulsar leaving a relic \PWN\ \citep{Blondin01}. This explanation has been confirmed by the detection of extended TeV emission matching the cocoon \citep{Aharonian06}, with the brightness peak offset from the pulsar, and of thermal X-ray emission suggesting mixing with the shocked ejecta \citep{LaMassa08}. However, an additional particle population, older and less energetic, is needed to explain the multiwavelength spectrum of Vela X \citep{deJager08}, complemented by  the recent GeV detections by {\em AGILE} \citep{Pellizzoni10} and {\em Fermi} \citep{Abdo10}. The angular resolution of IBIS/ISGRI on board the \integral\ observatory \citep{Winkler03} combined with its large field of view allows for the first time to address the problem of the full morphology of the hard X-ray nebula. 
Here we report on the \integral\ identification of extended hard X-ray emission from Vela.


\section{Observations and analysis}
\label{sec:2}

\subsection{IBIS/ISGRI imaging}
\label{sec:ISGRIimaging}
\begin{figure}[t]
\epsscale{1.0}
\includegraphics[keepaspectratio=true,width=0.98\textwidth,angle=0, clip=false]{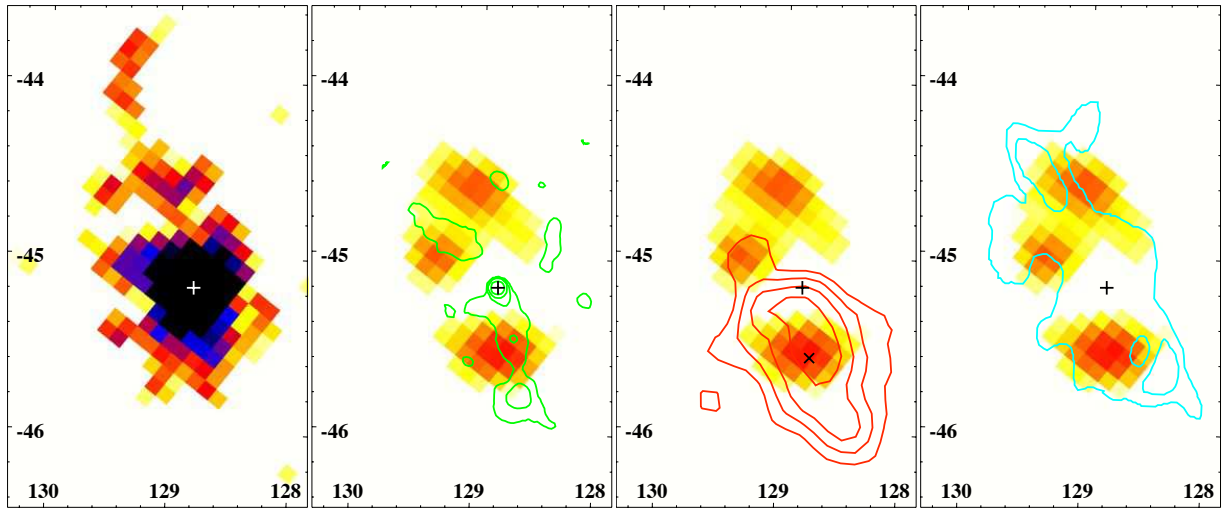}
\caption{IBIS/ISGRI significance map in the 18-40 keV range (celestial coordinates, J2000; north is up, east is left). The color scale is chosen to highlight the faint emission. First panel: significance map; other panels: significance map after subtraction of the point-like source, and smoothing with a 3 pixel ($\sigma$$\sim$7.5\arcmin) Gaussian kernel. Contours: {\em ROSAT}, 0.5-2 keV (second panel, green); \hess, VHE Gamma-rays above 1 TeV \citep[third panel, red,][]{Aharonian06}; Spacelab 2, 2.5-12 keV \citep[fourth panel, cyan,][]{Willmore92}. The cross indicates the pulsar position. The X point in the third panel marks the best-fit centre of gravity of the TeV emission. \label{fig:1}} 
\end{figure}

We analyzed all public \integral\ pointings within 12$^{\circ}$ from the Vela pulsar. We first analyzed the data from IBIS \citep{Ubertini03}, the coded mask imager on board \integral, and in particular of its low energy detector ISGRI  \citep[15 keV--1 MeV,][]{Lebrun03}. The IBIS/ISGRI data have been collected from 1976 pointings between 2003-03 and 2008-07, for a total exposure time of 5.6 Ms.
In the 18--40 keV mosaicked image (Fig. \ref{fig:1}), obtained with the Offline Scientific Analysis \citep[][]{Goldwurm03} software v.8, we found a 110$\sigma$ point-like source at the pulsar position. The point spread function (hereafter PSF) encompasses the pulsar, the \chandra\ PWN, and part of the fainter region.

An extended emission in the NE/SW direction is also visible in the image, spanning $\sim$50\arcmin\ on both sides. After subtraction of the point-like source by fitting it with a 2D Gaussian profile ($\sigma = 6.2$\arcmin), the NE side appears larger and more significant than the SW side, which is coincident with the \rosat\ and \hess\  cocoon (Fig. \ref{fig:1}, second and third panels). The extended emission also matches the one found by the Birmingham Spacelab 2 telescope in 2.5--10 keV (Fig. \ref{fig:1}, fourth panel). The individual pixels of the IBIS/ISGRI feature are at the $\sim$3--6 $\sigma$ significance level. Such a large cluster of low-significance pixels is not observed in the rest  of the image, and it is not reminiscent of IBIS/ISGRI coding noise. After smoothing, it is the only residual excess besides the known point sources. A further evidence of extended hard X-ray emission beyond the inner \PWN\ is provided by the spectral analysis.

\subsection{\integral\ spectral analysis}
\label{sec:INTEGRALspectrum}
\begin{figure}[t]
\epsscale{1.0}
\plotone{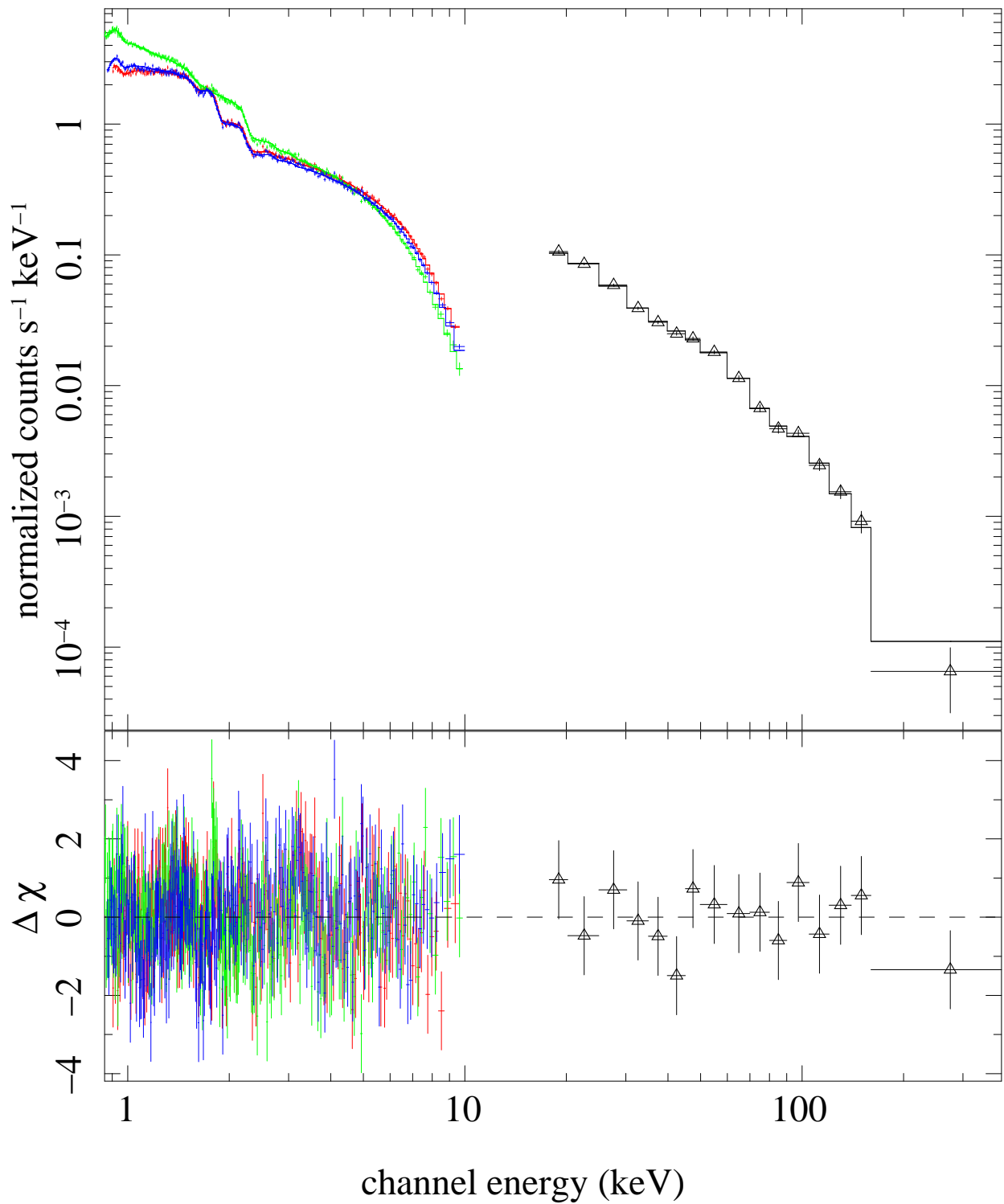}
\caption{Combined \suzaku/XIS0-1-3 (blue, green, and red) and IBIS/ISGRI (black) spectra of the Vela PWN within 6\arcmin\ from the pulsar fitted to the model with a Band component described in Sec. \ref{sec:combinedspectra}. The data have been graphically rebinned for clarity. The lower panel shows residuals from the best fit in units of 1$\sigma$. \label{fig:2}}
\end{figure}

We extracted the IBIS/ISGRI spectrum of the point-like source from mosaicked images in narrow energy bands between 18 and 400 keV (Fig. \ref{fig:2}). To account for the evolution of the instrument response, we produced an average of the responses weighted by the on-source exposures in their respective validity epochs. All the spectra in this work have been fitted using Xspec v.11.3 \citep{Arnaud96}, and the uncertainties are reported at the 90\% confidence level. A best fit with a single power law model yielded a photon index $\Gamma_{\mathrm{ISGRI}} = 2.00 \pm0.04$ and a flux $F = (4.76 \pm 0.09) \times10^{-11}$ \flux\ in the 20--40 keV range (\kir = 1.16/16 d.o.f.). This spectrum is $\sim$50 times higher than the phase-averaged one of the Vela pulsar at 20 keV \citep[Hermsen and Kuiper, priv. comm.;][]{Harding02}. Therefore, the IBIS/ISGRI emission is dominated by the nebula. 

We also analyzed the data from the \integral\ spectrometer SPI \citep[20 keV--8 MeV;][]{Vedrenne03} collected simultaneously to the IBIS data. SPI spectra have been extracted using the SPIROS package \citep{Skinner03} within the OSA analysis software. The SPI data are best fitted (\kir = 0.7/10 d.o.f.) by a power law model in the 20--300 keV range with photon index $\Gamma_{\mathrm{SPI}} = 2.15\pm0.15$, compatible within the errors with the IBIS/ISGRI one, but with a higher flux, $F = (9.1 \pm 0.6) \times 10^{-11}$ \flux\ in 20--40 keV, with respect to IBIS/ISGRI.  A joint fit to the IBIS/ISGRI and SPI spectra with a single power law yields a compatible photon index, but it requires a renormalization constant of 1.8 to recover the higher SPI flux. Such a discrepancy can not be accounted for by an intercalibration factor, which is in the 0.8--1.2 range for IBIS/ISGRI and SPI \citep[e.g.,][]{Jourdain08, Bouchet09}.

\begin{figure}[tb]
\epsscale{1.0}
\plotone{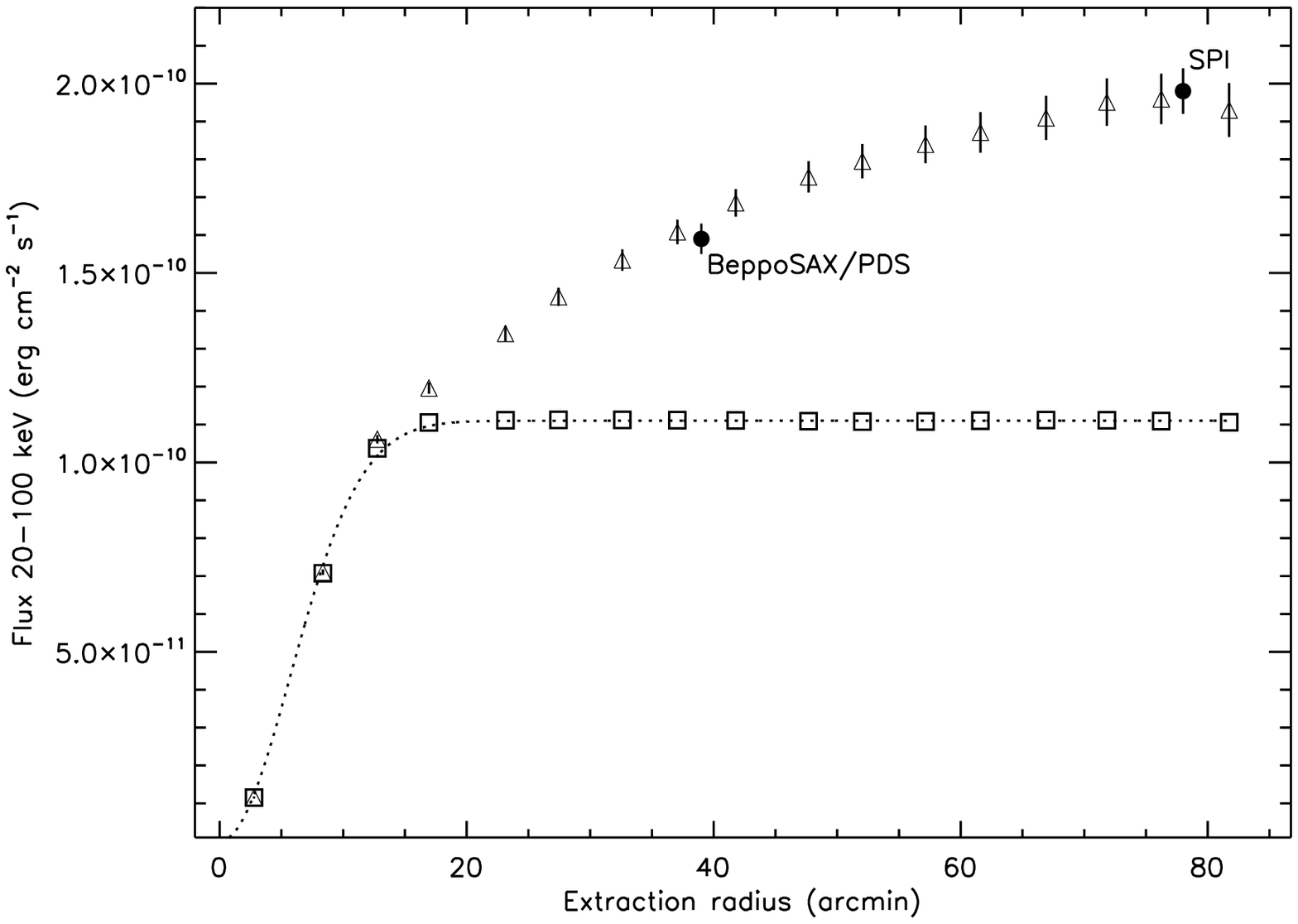}
\caption{IBIS/ISGRI flux in the 20--100 keV band integrated at varying extraction radii (open triangles), along with the \beppo/PDS and SPI fluxes in the same band (filled circles). In this case the integration radius is given by their PSF (39\arcmin, and 1.3\deg\ for \beppo/PDS, and SPI, respectively). The origin of the x-axis corresponds to the pulsar position. The expected (dotted line) and measured (the bright binary pulsar Vela X-1 rescaled, open squares) profiles for a truly point source are also shown. \label{fig:3}}
\end{figure}

The photon index measured by  \beppo/PDS in the same energy range \citep[$\Gamma_\mathrm{PDS} = 2.00 \pm 0.05$,][]{Mangano05} is consistent with both the spectral indices derived above, whereas the flux lies between the IBIS/ISGRI and SPI one. As shown in Fig. \ref{fig:3}, the IBIS/ISGRI, \beppo/PDS, and SPI fluxes correlate with the respective PSF radii (HWHM: 6\arcmin, 39\arcmin\footnote{As the PDS had no imaging capabilities, this is the HWHM of the instrumental angular response \citep{Frontera97}.}, and 1.3\deg), suggesting that each instrument samples a different portion of the nebula. Due to the coded mask deconvolution of IBIS/ISGRI, optimized for point sources, the reconstructed flux of an extended source of 60\arcmin\ radius is lower than the real one by a factor $\sim$50 \citep{Renaud06}. Therefore, a flux of $4.3 \times 10^{-11}$ \flux\ (the difference between the SPI and ISGRI fluxes) from such a source would be measured as low as $\sim$10$^{-12}$ \flux\ by IBIS/ISGRI, close to its sensitivity limit.

To support the hypothesis that an extended source is present in the IBIS/ISGRI data but diluted by the coded mask deconvolution, we refined the analysis following the method developed by \citet{Renaud06} for analyzing emission from extended sources.

We extracted the IBIS/ISGRI count rates from concentric circles centered on the pulsar with radii up to 80\arcmin, and converted them into flux by assuming a photon index as for the point-like source. This integrated flux as a function of the extraction radius does not reach a plateau after 15\arcmin, as expected for a point-like source, but slowly increases up to $\sim$60\arcmin\ (Fig. \ref{fig:3}). The integrated IBIS/ISGRI flux also recovers the \beppo/PDS and SPI fluxes at radii comparable with their PSFs. This confirms the detection of extended hard X-ray emission beyond the inner \PWN.

\subsection{IBIS/ISGRI and \suzaku/XIS combined spectrum}
\label{sec:combinedspectra}

The MECS and PDS instruments on board \beppo\ allowed to measure a break at energy $12.5 \pm 1.5$ keV \citep{Mangano05}. However, the different angular resolution of the two instruments required the authors to combine the PDS spectrum rescaled by an intercalibration factor with the MECS spectrum extracted from a 15\arcmin\ radius region. With a much smaller PSF, \integral\ IBIS/ISGRI can be combined with \suzaku/XIS on the same extraction radius.

\suzaku\ observed the Vela pulsar and PWN on 2006 July 10 and 11. Event files from version 2.0.6.13 of the \suzaku\ pipeline were used and spectra were extracted using {\sc XSELECT}. Response matrices and ancillary response files were generated for each XIS using {\sc XISRMFGEN} and {\sc XISSIMARFGEN} version 2007--05--14. The data of the XIS2 camera were not considered because of a more uncertain calibration. The effective exposure time for each XIS is about 60.3 ks. The source photons are extracted from a circular region with a radius of 6\arcmin\ to match the IBIS/ISGRI PSF. Background photons are extracted from blank sky observations within the same region as the source.

We modeled the pulsar contribution by introducing two black bodies corrected by the interstellar absorption, with fixed parameters as measured by \xmm\  \citep{Manzali07}. An absorbed non-equilibrium plasma emission model ({\sc VMEKAL}) is also included to account for the thermal supernova remnant; fixing the abundances as in \cite{LaMassa08}, we derived a temperature of $k_B T = 0.214^{+0.003}_{-0.005}$ keV. The XIS spectra show a bright power law component with photon index and flux (Table \ref{tab:1}) compatible with the MECS at the same radius \citep{Mangano05}. It is connected to the IBIS/ISGRI spectrum, confirming that the IBIS/ISGRI flux is due to the \PWN. However, a spectral break is required to simultaneously fit the XIS and the IBIS/ISGRI data. The break is located at higher energies ($27 \pm 3$ keV) than the one derived by \beppo. A simultaneous fit with a single power law is statistically rejected (\ki of 11252 for 6245 d.o.f.). 

We also refined the spectral fit by replacing the broken power law with a Band model, an empirical four-parameter model consisting of two power law components smoothly joined by an exponential roll-over \citep{Band93}. 
The fitted parameters are compatible with the ones derived with the broken power law model (Table \ref{tab:1}), for a comparable \ki. Notably, the two power laws intersect at $25 \pm 7$ keV, which corresponds to the break energy in the broken power-law model. In the next section we adopt the Band model, as a gradual transition should be more representative of the change of slope around a cooling break \citep[e.g.,][]{Kardashev62}.

\begin{deluxetable}{lcc}
\tablewidth{0pt}
\tablecaption{\footnotesize Best-fit spectral parameters.} 
\tablehead{ \colhead{} & \colhead{Broken power law} & \colhead{Band}}
\startdata
$\Gamma_1$ & 1.642$^{+0.005}_{-0.006}$ & 1.610$^{+0.012}_{-0.005}$ \\                   
$\Gamma_2$ & 2.07 $\pm$ 0.05 & 2.2$^{+0.1}_{-0.2}$ \\                   
$E_b$ (keV) & 27  $\pm$ 3 & \ldots \\
$E_0^{\ast}$ (keV) & \ldots & 114$^{+6}_{-16}$ \\                    
$F^{\dagger}$ & 1.11 $\pm$ 0.02 & 1.11 $\pm$ 0.06 \\ 
$\chi^2_{red}$ (d.o.f) & 1.03 (6235) & 1.03 (6235)
\enddata
\tablecomments{Best-fit spectral parameters of the combined \suzaku/XIS--IBIS/ISGRI data within 6\arcmin\ from the Vela pulsar. The uncertainties are at the 90\% confidence level. $^{\ast}$$E_0$ is the folding energy in the Band model. 
The two power laws intersect at $(\Gamma_2 - \Gamma_1) \, e^{-1} \, E_0 = 25 \pm 7$ keV, which corresponds to the break energy in the broken power-law model $E_b$. $^{\dagger}$Flux in the 20--100 keV energy band in units of 10$^{-10}$ \flux. \label{tab:1}}
\end{deluxetable}


\section{Discussion}
\label{sec:3}
Thanks to the deep \integral\ exposure, we were able to discover diffuse emission above 20 keV beyond the inner Vela \PWN. Such emission is resolved in two different regions: a southern hot-spot coincident with the Vela cocoon, notably with the peak of the TeV brightness profile measured by \hess, and a more extended northern emission, without any counterpart at other wavelengths and outside Vela X. However,  recent \suzaku/XIS observations showed non-thermal emission below 10 keV in a region located at the boundary of the IBIS northern emission \citep{Katsuda11}. We then compared the IBIS spectrum to the ones of \beppo/PDS and SPI, and explained the flux differences by their different PSFs, which sample different portions of the nebula, confirming a large extension ($\sim$1\deg\ radius) at hard X-rays. 

We also reported on the spectrum within 6\arcmin\ from the pulsar using the IBIS/ISGRI and the XIS telescopes. This is the first broad-band X-ray spectrum of the Vela \PWN\ taken within a region with the same angular extension below and above 10 keV. The change of slope around 25 keV ($\Delta \Gamma = 0.59 \pm 0.15$ for the Band model) is compatible with the standard value of 0.5 expected from a cooling break occurring in a continuously injected electron distribution affected by radiative losses. Indeed, the XIS and IBIS/ISGRI photon indices ($\sim$1.6 and $\sim$2.2) are compatible with the synchrotron spectrum of a shock-accelerated electron distribution in the uncooled and cooled regime, respectively  \citep[e.g.,][]{Chevalier00}. In this framework, the cooling energy is expected to decrease with time, that is at increasing integration radii from the pulsar. The lower break energy measured by  \beppo\ MECS and PDS on a larger angular extension may indicate a cooling break propagating along the flow. Indeed, the cooling energy in the cocoon is expected around 1 keV \citep{LaMassa08}. The similar photon indices found by IBIS, \beppo/PDS, and SPI suggest that the radiative losses above 20 keV already balance the injection rate within the region enclosed by the IBIS/ISGRI PSF. 

The measurement of a cooling energy at $\sim$25 keV in the photon spectrum allows us to set an upper limit on the time (residence time) spent by the particles in the region within 6\arcmin\ from the pulsar, corresponding to a distance of 0.5 $d_{290}$ pc. Accounting for synchrotron losses and inverse Compton losses in the Thomson regime,  the cooling frequency as a function of the residence time $t$ can be written
\begin{equation}
\label{eq:coolingfrequency}
\nu_c  (t) =  \frac{81 \, m_e^5 \, c^9}{32 \pi \, e^7 \, (1 + U_{ph}/U_B)^2 \, B^3  \, t^2},
\end{equation}
where $U_{ph}$ and $U_B$ are the photon field and magnetic energy densities, respectively. Eq. (\ref{eq:coolingfrequency}) coincides with the cooling frequency calculated assuming only synchrotron losses \citep[e.g.,][]{Chevalier00} for $B \gg \sqrt{8 \pi U_{ph}}$. Solving it for $t$, the residence time has a maximum occurring for a magnetic field $B = \sqrt{8 \pi U_{ph} /3}$ independently from the cooling frequency. Such a magnetic field amounts to 1.9 $\mu$G when the target photons are provided by the Cosmic Microwave Background radiation (CMB, $U_{ph} = 0.26$ eV cm$^{-3}$). For a cooling energy of 25 keV ($\nu_c = 6 \times 10^{18}$ Hz), the residence time of electrons is 1650 yr in this case, and shorter for any different intensity of the magnetic field and any additional photon field. Therefore, the electrons radiating in X-rays in the considered region can not be older than 1650 yr. We conclude that those injected before have flowed out of the region within 6\arcmin\ from the pulsar. This requires a moderate average velocity ($> 300 \, d_{290}$ km/s).

\begin{figure}
\epsscale{1.0} 
\plotone{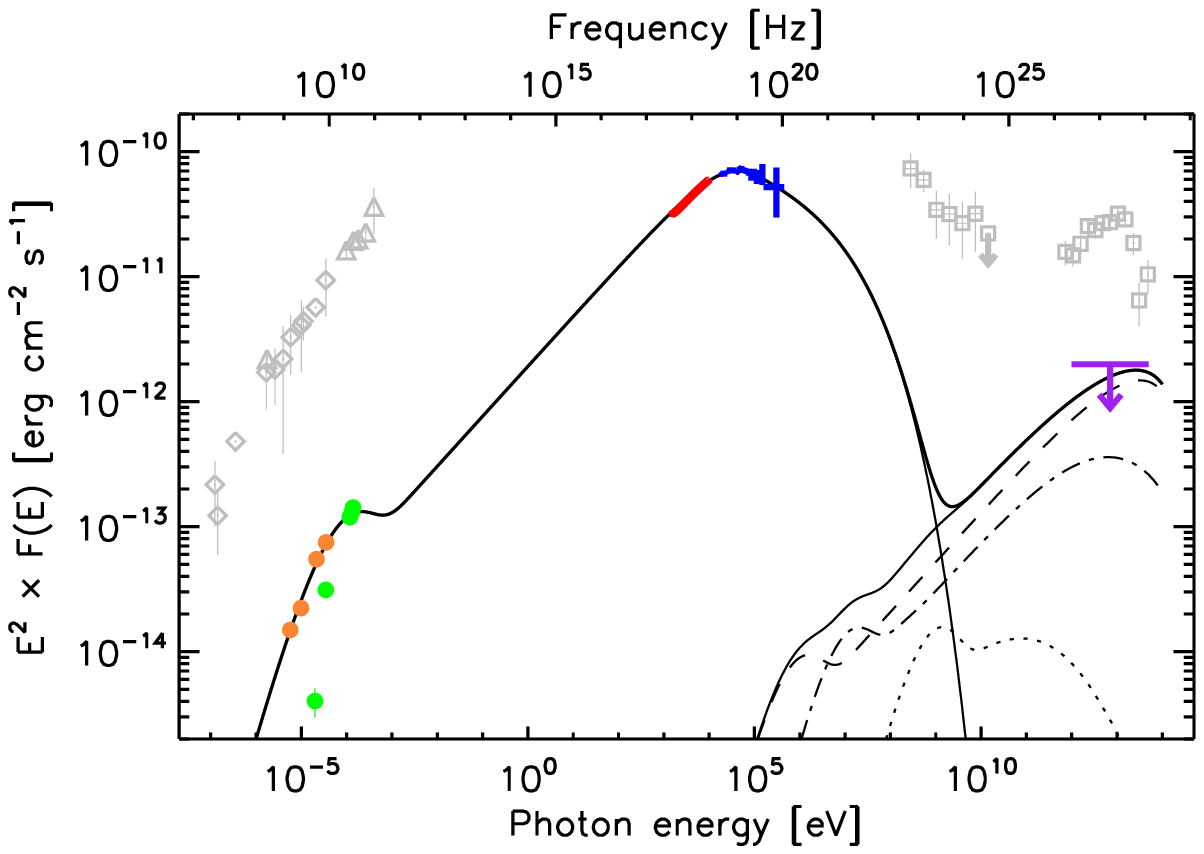}
\caption{S.E.D. of the Vela \PWN\ emission within 6\arcmin\ from the pulsar, fitted with the model described in the text. The \suzaku\ XIS and \integral\ IBIS/ISGRI spectra are shown in red and blue, respectively. The radio fluxes of the inner Vela \PWN\ are shown \citep[][orange and green circles respectively]{Dodson03b, Hales04}. The upper limit (99.9\%) on the integral flux above 1 TeV within 6\arcmin\ from the pulsar is also shown \citep[purple arrow, assuming a photon index of 2;][]{Aharonian06}. The measurements of the large-scale \PWN\ are reported in grey for comparison: Vela X in radio \citep{Alvarez01, Abdo10}, at GeV energies \citep{Abdo10}, and the TeV cocoon \citep{Aharonian06}. The total (synchrotron and IC) model spectrum is indicated with a thick (thin) solid line. The IC emission is computed taking into account the CMB (dashed line), dust (dot-dashed line), and star-light (dotted line). The magnetic field is 10 $\mu$G. \label{fig:4}}
\end{figure}

We explored the hypothesis that particles of all energies remain in this region for the residence time which yields a cooling break at 25 keV for a given $B$. This is done by means of a time-dependent one-zone model of the spectral energy distribution (S.E.D., see Fig. \ref{fig:4}). The injection spectrum is composed by a relativistic Maxwellian \citep[][]{Sironi09a} and a cut-off power law with index 2.2. The radiative losses depend on a magnetic field of constant intensity $B$ and a target photon field taken at the Vela position according to \citet{Moskalenko06}. The cut-off energy is fixed by $B$ through the condition that the acceleration rate equals the cooling rate \citep{deJager96}. The shape of the distribution is constant, whereas the normalization is proportional to the pulsar spin-down power:
\begin{equation}
\dot{N} = \frac{\eta \, \dot{E}}{\Gamma_W \, m_e c^2},
\end{equation}
where $\eta$ is the fraction of $\dot{E}$ converted into the wind energy, and $\Gamma_W$ is the average Lorentz factor fixed by the energy conservation. Such an injection rate decreases in time following $\dot{E}(t) = \dot{E}_0  \, ( 1 + t/t_{dec})^{-\beta}$, where $\dot{E}_0$ is the initial spin-down power, $t_{dec}$ the spin-down time-scale, and $\beta = {(n+1)/(n-1)}$ for a braking index $n$ \citep[][]{Pacini73}.

The S.E.D. in Fig. \ref{fig:4} is reproduced by this simple model for a range of values of $B$. The upper limit on the integral flux above 1 TeV implies a firm lower limit on the magnetic field, $B > 10 \, \mu$G, higher than the one estimated in the cocoon \citep[][]{deJager08}. This result depends on the relative intensity of the synchrotron and Compton peaks, which is not affected by the conversion efficiency $\eta$ (1.3\% for $B = 10 \, \mu$G). It is also nearly irrespective of $t_{dec}$ and $n$: given the short evolution time, a manifold break around the cooling energy due to the pulsar spin-down \citep[][]{Gelfand09} is not evident\footnote{We tested  $t_{dec}$ between 100 yr and 1000 yr, and both $n=3$ (dipolar rotator model) and $n = 1.6$ \citep[measured,][]{Dodson07}. $\dot{E}_0$ has been chosen to yield $\dot{E} = 6.9 \times 10^{36}$ erg s$^{-1}$ at the present time.}. The extrapolation of the X-ray spectrum at lower energies falls squarely on the measured radio fluxes of the compact \PWN. A fine tuning is obtained adopting a relativistic Maxwellian with a temperature of 3 GeV for $B = 10 \, \mu$G; alternatively, a low-energy break in the injected electron population is required at GeV energies, as found in young \PWNe\ \citep[e.g.,][]{Bucciantini11}.  Additional measurements between radio and X-ray frequencies are needed to exclude a more complex spectrum \citep[e.g.,][]{Slane08}. 

The fact that diffuse hard X-ray emission is detected beyond 6\arcmin\ from the pulsar in the northern direction strengthens the hypothesis of particle leakage. Unlike the southern hot-spot, which may simply be the high-energy counterpart of the X-ray cocoon, the northern emission may be generated by particles injected after the transit of the supernova reverse shock \citep[$\sim$3 kyr ago, ][]{Blondin01}. The original \PWN\ is then set apart from the pulsar, becoming relic, while the latter forms a new \PWN\ in subsonic expansion inside the supernova remnant \citep{vanderswaluw04}. 
To understand the nature of the northern emission it will be crucial to extend the X-ray mapping of the Vela PWN with \xmm\ and \suzaku\ and with the forthcoming focusing telescopes at higher energy X-rays (Nu-STAR and ASTRO-H).

\acknowledgments
Based on observations with \integral, an ESA mission with instruments and science data centre funded by ESA member states, Czech Republic, and Poland, and with the participation of Russia and the USA. ISGRI has been realized and maintained in flight by CEA-Saclay/Irfu with the support of CNES. FM, IC, SS, and JAZH acknowledge support from CNES through CNRS. GP acknowledges support via an EU Marie Curie Fellowship under contract no. FP7-PEOPLE-2009-IEF-254279. We are grateful to Wim Hermsen and Lucien Kuiper for useful discussions.




{\it Facilities:} \facility{INTEGRAL (IBIS/ISGRI, SPI)}, \facility{Suzaku (XIS)}.

\bibliographystyle{apj}

\end{document}